\documentclass[usenatbib]{mn2e}
\usepackage{amsfonts}
\usepackage{amssymb}
\usepackage{amsmath}
\usepackage{graphicx}
\usepackage{fixltx2e}
\usepackage{bm}

\newcommand{\etal}{{et al.}~}

\newcommand{\eg}{{e.g.~}}
\newcommand{\ie}{{i.e.~}}

\def \ltsima{$\; \buildrel < \over \sim \;$}
\def \simlt{\lower.5ex\hbox{\ltsima}}            
\def \gtsima{$\; \buildrel > \over \sim \;$}
\def \gtsima{\mbox{$\; \buildrel > \over \sim \;$}}
\def \simgt{\lower.5ex\hbox{\gtsima}}            

\newcommand{\be}{\begin{equation}}
\newcommand{\ee}{\end{equation}}
\newcommand{\ba}{\begin{eqnarray}}
\newcommand{\ea}{\end{eqnarray}}

\def\pp1{{\prime}}
\def\pp2{{\prime\prime}}

\def\2D{{\rm 2D}}

\def\1Loop{{\rm 1Loop}}

\def\Msol{h^{-1}M_{\odot}}

\def\kpc{\, h^{-1}{\rm kpc}}

\def\keV{{\rm keV}}

\def\la{\mathrel{\mathpalette\fun <}}
\def\ga{\mathrel{\mathpalette\fun >}}
\def\fun#1#2{\lower3.6pt\vbox{\baselineskip0pt\lineskip.9pt
        \ialign{$\mathsurround=0pt#1\hfill##\hfil$\crcr#2\crcr\sim\crcr}}}

  \title[The cosmic web of WDM versus CDM dwarfs]
  {The same with less: The cosmic web of warm versus cold dark matter dwarf galaxies}
\author[Reed \etal] 
{\parbox{\textwidth}
{
Darren~S.~Reed$^{1*}$
Aurel Schneider$^{2,3}$,
Robert~E.~Smith$^{3}$,\\
Doug Potter$^{2}$,
Joachim Stadel$^{2}$ 
\&
Ben Moore$^{2}$ \\
}
\\
{$^1$Institut de Ci\`encies de l'Espai (IEEC-CSIC), 08193 Bellaterra (Barcelona), Spain}\\
{$^2$Insitute for Computational Science, Univ. of Z\"{u}rich, Winterthurerstrasse 190, CH-8057 Z\"{u}rich, Switzerland}\\
{$^3$Astronomy Centre, Department of Physics and Astronomy, University of Sussex, BN1 9QH, Brighton, United Kingdom}\\
}

\pagerange{\pageref{firstpage}--\pageref{lastpage}}

\begin{document}

\maketitle

\label{firstpage}


\begin{abstract}
  We explore fundamental properties of the distribution of low mass
  dark matter halos within the cosmic web using warm dark matter (WDM)
  and cold dark matter (CDM) cosmological simulations.  Using self
  abundance-matched mock galaxy catalogs, we show that the
  distribution of dwarf galaxies in a WDM universe, wherein low mass
  halo formation is heavily suppressed, is nearly indistinguishable to
  that of a CDM universe whose low mass halos are not seen because
  galaxy formation is suppressed below some threshold halo mass.
  However, if the scatter between dwarf galaxy luminosity and halo
  properties is large enough, low mass CDM halos would sometimes host
  relatively bright galaxies thereby populating CDM voids with the
  occasional isolated galaxy and reducing the numbers of completely
  empty voids.  Otherwise, without high mass to light scatter, all
  mock galaxy clustering statistics that we consider--the
  auto-correlation function, the numbers and radial profiles of
  satellites, the numbers of isolated galaxies, and the PDF of small
  voids--are nearly identical in CDM and WDM.  WDM voids are neither
  larger nor emptier than CDM voids, when constructed from
  abundance-matched halo catalogs.  It is thus a challenge to
  determine whether the CDM problem of the over-abundance of small
  halos with respect to the number density of observed dwarf galaxies
  has a cosmological solution or an astrophysical solution.  However,
  some clues about the dark matter particle and the scatter between
  the properties of dwarf galaxies and their dark matter halo hosts
  might be found in the cosmic web of galaxies in future surveys of
  the local volume.

\end{abstract}




\section{introduction}
\label{sec-introduction}

\let\thefootnote\relax\footnotetext{*~email:~reed@physik.uzh.ch} 

In Warm Dark Matter (WDM) cosmological models, small scale power is
suppressed below the mass-scale of dwarf galaxy halos due to
relativistic free-streaming when particles freeze-out from the
matter-radiation field.  In contrast, Cold Dark Matter (CDM) particles
are ``cold'' (non-relativistic) at freeze-out, so CDM small scale
structure is preserved down to ``micro-halo'' scales, \ie
$10^{-6}\Msol$ plus or minus several orders of magnitude
\citep[\eg][]{Hofmannetal2001,Bertoneetal2005,Greenetal2005,Diemandetal2005c,Profumoetal2006,Bringmann2009}.
CDM cosmology with a cosmological constant ($\Lambda$CDM) has been
successful at reproducing a number of large-scale observations,
including the cosmic microwave background \citep[CMB,
][]{Planckcosmology2014short}, the large-scale clustering of galaxies
\citep{Percivaletal2010short,Andersonetal2012}, and the mass function
of clusters of galaxies \citep{Allenetal2011,Mantzetal2015}.  However,
none of these observations directly sample the matter power spectrum
or halo mass function on mass-scales below that of bright ($\sim L_*$)
galaxies, which means that CDM and WDM are both consistent with
large-scale cosmological probes.  Moreover, several measurements of
structure on small scales are difficult to explain with CDM and have
been cited as possible evidence for WDM.  A number of possible
physical mechanisms for producing WDM have been proposed
\citep{Colombietal1996,Kawasakietal1997,Boyarskyetal2009b}.

Among the observations that implicate a warm particle are the reduced
number of satellite galaxies in the Milky Way and M31 relative to the
number of CDM satellites from simulations
\citep{Mooreetal1999b,Klypinetal1999}.  The over-abundance of small
halos relative to galaxy numbers is not limited to the local group but
extends to the flat field optical and HI circular velocity functions
and the faint galaxy luminosity functions relative to the steep low
mass halo mass function or circular velocity function
\citep[\eg][]{Blantonetal2001,Zavalaetal2009,Trujillo-Gomezetal2011,Papastergisetal2015,Schneideretal2014,Klypinetal2014}.
A thermal relic of $2~keV$ is able reduce the numbers of WDM
satellites to the number of observed satellites in cosmological
simulations \citep[\eg][]{PolisenskyRicotti2011}.  The lower
concentrations of WDM halos \citep{Schneideretal2012, Schneider2015} corresponds to
lower densities in the inner regions (\ie where there are stars),
which brings the satellite halo circular velocities into better
agreement with Local Group galaxy rotation curves
\citep{Lovelletal2012}, pushing the ``missing satellite'' issue to
lower masses where star formation is more easily suppressed.
Inefficient star and galaxy formation within low mass halos has also
been proposed as a solution within CDM cosmology
\citep[\eg][]{Bensonetal2003}.  Hence either a {\it cosmological }
solution (warm dark matter) or an {\it astrophysical } solution
(baryon physics) have the potential to explain some of the small-scale
CDM problems.

Both classes of proposed solutions have unsolved issues.  Recent
inferences of the matter power spectrum on small scales from
Lyman-alpha (Ly-$\alpha$) forest lines in quasar spectra disfavor a
dark matter particle warm enough to reduce sufficiently the numbers of
low mass halos for agreement with the numbers of dwarf galaxies
\citep{Vieletal2013}.  However, the astrophysical solution to the CDM
small-scale structure problem also presents a challenge because it
requires that star formation in the largest Milky Way satellites be
quenched while still allowing some galaxies to form in much smaller
halos, as inferred from halo rotation curves and stellar kinematics --
the ``too big to fail'' problem \citep{Boylan-Kolchinetal2011}.  This
implies a need for large scatter in halo luminosity to mass if we have
a CDM Universe.

Regardless of whether WDM is allowed by the Ly-$\alpha$ forest to be
warm enough to prevent the over production of small structures in CDM,
or whether baryon physics is able to hide them, there remain no
constraints against WDM that is a bit less warm.  ``Lukewarm'' DM
particles of 4keV or cooler remain viable candidates
\citep{Vieletal2013}.  Moreover, it is important to search for
independent local constraints on the DM particle.  For these reasons,
we will explore whether a warm dark matter particle might leave an
imprint in the cosmic web of halos beyond the main effect of
suppressing low mass halo numbers.

If one understood galaxy formation and associated baryonic effects on
dark matter well enough to accurately map galaxy properties to halo
mass, one could probe small-scale initial power by directly measuring
the halo mass function at dwarf galaxy scales.  However, the poorly
understood effects of baryon physics on inner halo profiles combined
with the mostly unknown scatter between halo mass and dwarf galaxy
luminosities and other properties make this difficult.  In particular,
gravitational coupling of baryons to dark matter via gas cooling
associated with star formation and gas ejection from stellar feedback
has been shown in simulations to reduce central dark matter densities
\citep{Mashchenkoetal2006,PontzenGovernato2012,Governatoetal2012}.
This baryon feedback could perhaps transform CDM halo central density
cusps into shallower cores, which would be a better apparent match to
observed galaxies profiles \citep[][and others]{Mooreetal1999c}.
However, any such baryon gravitational coupling to the dark matter
also makes it difficult to infer dark halo masses from galaxy
kinematics, and the ``true'' (\ie unaltered by baryons) low mass halo
circular velocity function may be much steeper than inferred from
observations.  It is thus valuable to seek additional tests that might
be able to distinguish between WDM and CDM cosmology.

Recent work has revealed subtle differences in the distribution of
dwarf galaxies that might be exploited with galaxy surveys.  The sizes
of voids, defined by the distribution of halos, have often been noted
to be larger and emptier in WDM than CDM simulations because there are
more low mass halos in CDM
\citep{TikhonovKlypin2009,Tikhonovetal2009}.  The striking emptiness
of voids in galaxy surveys has sometimes been cited as a problem for
CDM \citep{Peebles2001}.  However, since voids are delineated by
galaxies (lying in halos), the sizes and emptiness of voids is highly
sensitive to the number abundance of galaxies in a survey, and thus a
comparison between observed and CDM voids requires knowledge of how
galaxies populate halos.  A low efficiency of galaxy formation in low
mass halos is able to create large and empty CDM voids, in agreement
with those found in the local volume, by the effect on galaxy number
density \citep{TinkerConroy2009}, though it violates the ``too big to
fail'' test in that some very low $V_c$ galaxies are observed below
the galaxy formation cutoff
\citep{Tikhonovetal2009,Papastergisetal2015}.

Other work has shown that halos in a very warm DM cosmology should
have somewhat stronger large-scale clustering
\citep{SmithMarkovic2011,Dunstanetal2011,Schneideretal2012} because
they are biased to lie in denser regions according to the
peak-background split halo clustering model
\citep{MoWhite1996,ShethTormen1999}.  Our work expands upon these
studies by exploring some of the fundamental clustering properties of
WDM halos relative to CDM halos.  We focus on the dwarf galaxy halo
mass range and include both field halos and satellite halos (subhalos)
within larger galaxy halos.

We use numerical simulations, described in \S~\ref{sec-simulations},
to address the potential of low mass halo and mock galaxy clustering
to differentiate between WDM and CDM cosmology.  In
\S~\ref{sec-results}, we show that WDM halo and galaxy clustering
strength is very similar to that of CDM in our mock catalogs.  We then
demonstrate in \S~\ref{sec-voidsneighbors} that other clustering
measures such as the volume fraction occupied by voids and the
probability distribution function (PDF) of nearest neighbor distances
are also nearly the same in the two cosmologies.  While this makes it
difficult to use the cosmic web of galaxies to distinguish WDM and
CDM, we discuss in \S~\ref{sec-galaxies} that some differences arise
in small void statistics of cold and warm cosmology, if there is large
scatter between galaxy luminosity and halo mass.  Thus, there is the
potential for observable signals of the dark matter particle type to
be measured in the properties of small voids.  Finally, in
\S~\ref{sec-strangelove}, we discuss numerical issues that can affect
WDM simulations and show evidence that any related systematic errors
are not significant in our mock catalogs and do not affect our
conclusions.

\section{the simulations}
\label{sec-simulations}

The initial distribution of particles is created using relatively
standard techniques as discussed in
\citep[\eg][]{Scoccimarro1998,Crocceetal2006,Prunetetal2008,Reedetal2013}.  
We make use of a slightly modified version of the publicly available
code {\tt 2LPT}\footnote{http://cosmo.nyu.edu/roman/2LPT}, introduced
by \citet{Crocceetal2006} to reduce early numerical errors, 
``transients'', caused by the fact that the simulation must be initialized at some finite redshift when the density field is no longer accurately described by linear perturbations.
Simulations are initialized at $z_i=100$, approximately 10
expansions before the formation of the first generations of halos at
$z \sim 10$.  This start is early enough to model accurately
the formation of early CDM halos, and it should be late enough
that initial cosmological power is sufficiently large such that real cosmological forces dominate over spurious forces.
Moreover, because the WDM and CDM simulations use the same
initialization epoch and identical simulation parameters, 
the effects of any potential numerical inaccuracies can be minimized in 
comparisons between the two cosmologies.

Two twin 25 Mpc volumes are simulated, one with a $\Lambda$CDM
cosmology and one with a $\Lambda$WDM cosmology.  We use the
\citet{EisensteinHu1998} transfer function to set the CDM initial
conditions power.  The $\Lambda$WDM assumes a $2~keV$ thermal relic
warm particle, and we include some results from warmer simulations of
0.1 and $0.5~keV$ from \citep{Schneideretal2013}.  By focusing on the
$2~keV$ thermal relic, which as we discussed in the previous section,
appears to be too warm for Lyman-$\alpha$ forest data
\citep{Vieletal2013}, we maximize any differences between WDM and CDM
-- the goal of this work being to determine whether any such
differences might be detectable.  A small scale power suppression
using the prescription of \citet{Vieletal2005} is applied to the CDM
transfer function to generate the WDM transfer function.  The CDM and
WDM particle distribution has been constructed with the same
statistical realisation to minimize the effects of finite sampling of
large-scale waves when comparing the two simulations.  Initial WDM
thermal motions are neglected because they are expected to be
insignificant relative to the halo kinematics by low redshift
\citep{Bodeetal2001}.  Each volume is a periodic cube of $1024^3$
equal mass particles, for a particle mass of $3.9\times10^5\Msol$.  A
``Hann filter'', which reduces initial anisotropies in the density
field \citep{Bertschinger2001} but also suppresses initial small-scale
power \citep[see discussion in ][]{Reedetal2013}, is not used.  While
the simulation volume is small enough to suffer finite volume errors
that suppress massive halo numbers, our focus on the scale of dwarf
galaxies and the nature of our comparative study between WDM and CDM
allow a relatively small box to be used, which saves computational
resources.

Simulations are evolved using the particle gravity tree-code {\tt
 PKDGRAV}, an early version of which is described in
\citet{Stadel2001} and \citet{Wadsleyetal2004}, with numerical
accuracy parameters consistent with converged values in
\citet{Reedetal2013}.  Force resolution is set by the comoving
softening length of $\epsilon=0.5$kpc.  The adaptive time-step length
criterion, $\eta=\sqrt(\epsilon/a)$, where $a$ is the acceleration
acting on each particle, is set to $\eta=0.2$.  Medium and long range
force accuracy is governed by the tree opening angle, $\Theta = 0.7$.

\subsection{Warming the initial power spectrum}
\label{sec-wdmtheory}

We briefly review the technique we use to transform the CDM initial
fluctuation spectrum into a WDM spectrum, noting the important mass
and length scales.

Several fitting formulas for the WDM density transfer function have
been proposed \citep{Bardeenetal1986,Bodeetal2001}; we adopt the
formula in \citet{Vieletal2005}:
\be T_{\rm
  WDM}(k)=\left[\frac{P_{\rm lin}^{\rm WDM}}{P_{\rm lin}^{\rm
      CDM}}\right]^{1/2}=\left[1+(\alpha k)^{2 \mu}\right]^{-5/\mu},
\label{TFwdm}
\ee 
with $\mu=1.12$, and the {\it effective} free-streaming length, below which initial density perturbations are insignificant, is
\be 
\alpha = 0.049 \left[\frac{m_{\rm WDM}}{\keV}\right]^{-1.11}
\left[\frac{\Omega_{\rm
      WDM}}{0.25}\right]^{0.11}\left[\frac{h}{0.7}\right]^{1.22}\rm
Mpc/h.
\ee
Here, we assume a fully thermalized WDM particle. 

Another important scale is the `half-mode' scale, the scale at which
the amplitude of the WDM transfer function is reduced to 1/2 the CDM
value.  The half-mode length scale is given by:
\be \lambda_{\rm hm}=2\pi\lambda^{\rm eff}_{\rm
  fs}\left(2^{\mu/5}-1\right)^{-1/2\mu} \approx 13.93 \alpha\ .  \ee
\citep[see \eg][]{Schneideretal2012}.
The corresponding half-mode mass scale is then
\be M_{\rm hm}=\frac{4\pi}{3}\overline{\rho}\left(\frac{\lambda_{\rm
      hm}}{2}\right)^3 \approx 2.7\times 10^3 M_{\rm fs}\ ,  
\label{eqn-mhm}
\ee
or  $1.25 \times 10^{9} h^{-1} \Msol$ for our $2~keV$ relic.
$M_{\rm hm}$ can be thought of as the approximate halo mass below
which the WDM mass function diverges by a factor of a few or more from
CDM.


\subsection{Halos and simple mock galaxies}
\label{sec-halos}

In order to identify halos that can host galaxies, we use the Amiga
Halo Finder \cite[{\small AHF}][]{Gilletal2004,KnollmannKnebe2009}.
AHF finds self-bound field, central and satellite halos (\ie halos and
subhalos).  Field halos are those that satisfy a spherical
over-density criterion corresponding to virialized objects and may
contain multiple subhalos defined by identifying local density maxima
and the matter bound to them.  We construct a simple mock galaxy
catalog by allowing every halo, field or satellite, to host one galaxy
at its center.  Mock galaxies are selected by the circular velocity
($V_c$) at the peak of the rotation curve.  Because $V_{c}$ is less
affected by tidal stripping than mass, it is likely a better indicator
of pre-infall halo mass for the case of satellites, and thus is likely
to better correlate to galaxy stellar mass, according to generally
accepted models of galaxy formation.  $V_c$-selected halo and subhalo
catalogs have had success at matching the numbers and distribution of
galaxies larger than ${\rm 80 km~s^{-1}}$
\citep{Trujillo-Gomezetal2011}.  Mass-selected AHF and
Friends-of-friends \citep[FoF,][]{Davisetal1985} halos, which
correspond more closely to virialized halos, are also considered
separately in \S\ref{sec-strangelove}.

Mock galaxy numbers in CDM and WDM are matched by abundance, initially
with a one-to-one relation between galaxy luminosity and dark halo
$V_c$.  It is well known that absolute numbers of dwarf halos are
suppressed in WDM, and we should make the conservative assumption that
stellar kinematics do not necessarily reflect halo mass.  Thus, a fair
comparison between the CDM and WDM cosmic web requires that the CDM
and WDM mock galaxy catalogs be matched by galaxy number to reflect
the possibility that we may live in a CDM universe whose low mass
halos do not become galaxies.  We also consider the effect of scatter
between luminosity and $V_c$ on our mock galaxy distribution.

\begin{figure}
  \includegraphics[type=pdf,ext=.pdf,read=.pdf,width=.47\textwidth]
{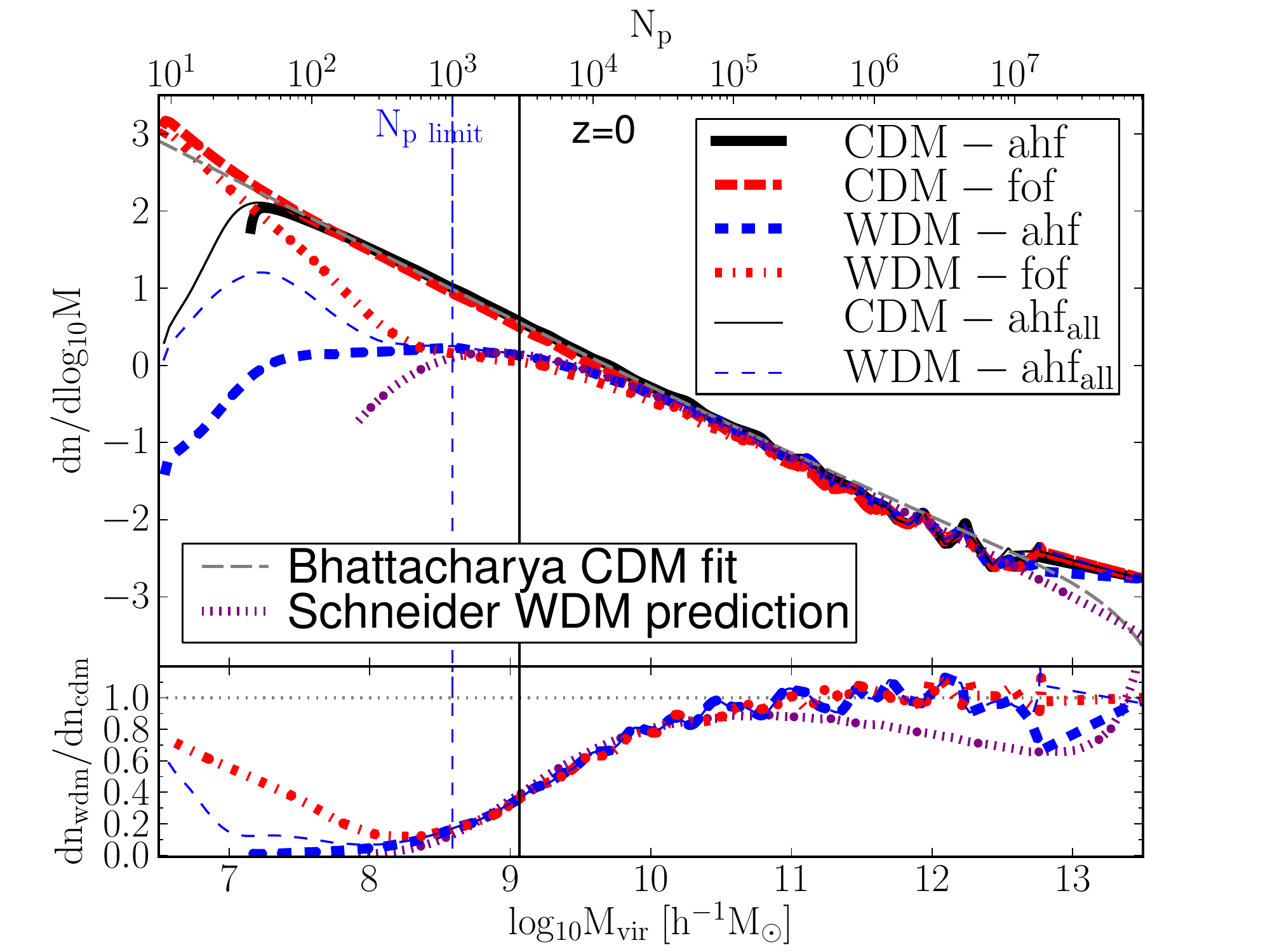}
  \includegraphics[type=pdf,ext=.pdf,read=.pdf,width=.47\textwidth]
{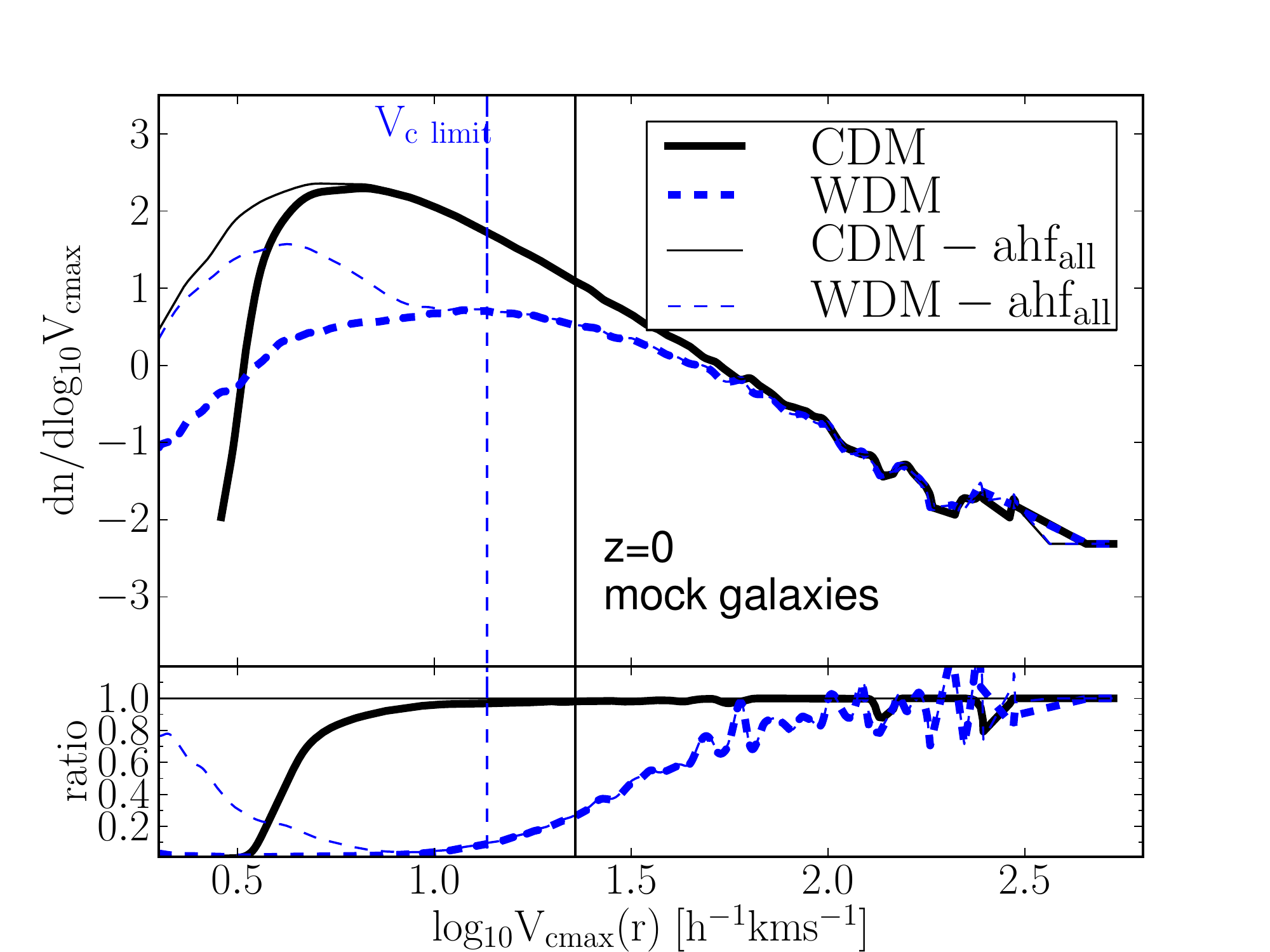}
  \caption{The halo mass function (top) and circular velocity function
    (bottom) for the WDM and the CDM runs for FoF and AHF halos.
    Dashed vertical lines denote the minimum effective halo mass (or
    $V_c$) included in our mock galaxy catalogs, chosen to exclude the
    mass range of spurious WDM halo formation, indicated by the low
    mass upturn.  The WDM AHF halos, which include satellites, have a
    similar mass function and a similar minimum resolved mass as FoF
    halos.  Thick AHF curves have been purged of spurious halos based
    on the sphericity of the initial pre-collapse halo Lagrangian
    volume, based on \citet{Lovelletal2014}; thin AHF curves include
    all AHF halos.}
  \label{fig-mf}
 \end{figure}

When constructing our mock galaxy catalog, we set the minimum $V_c$ to
match the abundance of WDM halos with more than 1000 particles ($5.5
\times 10^{8}\Msol$), approximately double the mass where spurious WDM
halos, which form from the artificial fragmentation of WDM filaments
\citep{WangWhite2007}, begin to become important; see further
discussion of numerical issues related to WDM in particular in
\S\ref{sec-strangelove}.  This conservative choice for minimum $V_c$
results in approximately $10^4$ mock galaxies with ${\rm V_c > 13.7 km
  ~s^{-1}}$ (WDM) and ${\rm V_c > 22.9 km ~s^{-1}}$ (CDM).  In
addition to the halo $V_c$ resolution limit, we impose the restriction
that the halo particles be formed from an initial Lagrangian volume
that does not deviate significantly from spherical, an indicator of
artificial structure \citep{Lovelletal2014}.  Fig. \ref{fig-mf} shows
the halo catalogs and the effect of the removal of spurious halos.
Even with our conservative resolution limits, we sample well into the
mass and $V_c$ range where WDM halo numbers are reduced, so we are
able to examine any important differences caused by WDM-like
suppression of structure formation.  Our minimum halo $V_c$ for
inclusion in the catalog is large enough that artificial halo numbers
are insignificant.  We thus believe that the amount of contamination
from artificial structure is small and does not significantly impact
our results.

\section{the similarity of WDM and CDM clustering}
\label{sec-results}

We first consider the pair auto-correlation function, $\xi$, as a
measure of clustering strength.  We estimate $\xi$ via a histogram of
halo pairs binned by radius:
\begin{equation}
\xi(r)=N_{\rm pairs}(r)/N_{\rm pairs, random}(r)-1,
\end{equation}
where ${\rm N_{pairs, random}}$ is estimated from the number density.
Note that $\xi$ is a essentially the Fourier transform of the halo
power spectrum \citep[\eg][]{Peebles1980}.  To determine $\xi$ of the
mass, we take a random subsample of $10^5$ simulation particles,
enough that statistical errors are much smaller than for halo pairs,
and measure the same pair statistics as for halos.

\begin{figure}
  \includegraphics[type=pdf,ext=.pdf,read=.pdf,width=.52\textwidth]{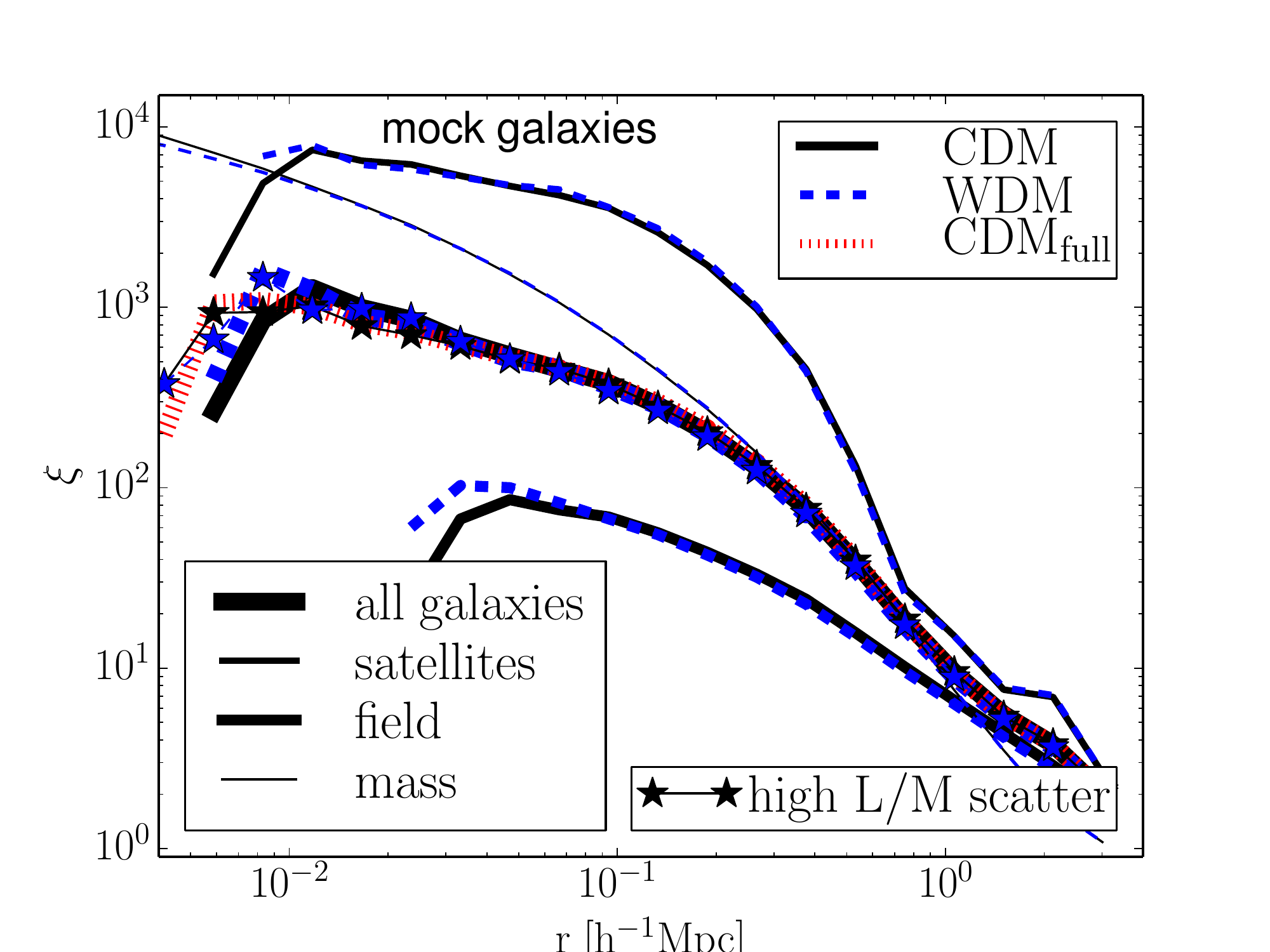}
  \caption{ The correlation function of our mock galaxy catalogs
    selected by halo circular velocity (${\rm V_c > 13.7 km ~s^{-1}}$
    for WDM and ${\rm V_c > 22.9 km ~s^{-1}}$ for CDM) so that the
    abundance matches in the two cosmologies.  The clustering in WDM
    and CDM is nearly indistinguishable.  The CDM$_{\rm full}$ sample
    uses the same $V_c$ threshold as the WDM sample, showing that the
    selection criteria has little effect on CDM halo clustering.  The
    stars show a mock galaxy catalog that includes a large scatter in
    mass to light ratio, which could be present in the dwarf galaxy
    population, described in \S\ref{sec-voidsneighbors}; this scatter
    has little effect on $\xi$.  }
  \label{fig-corr}
 \end{figure}

Fig. \ref{fig-corr} compares the CDM and WDM mock galaxy catalogs,
matched to each other by abundance.  The differences in clustering
between WDM and CDM mock galaxies are small, consistent with zero,
given the uncertainties, and noting that shot noise is large at the
smallest separations.  One can estimate from the bin to bin scatter in
the correlation function that the shot noise uncertainty is $\simgt 10
\%$ on small scales.  The sample variance is larger than the shot
noise because of our small boxes, but is not important for the CDM
versus WDM comparison.  Again, a $V_{c}$-selected halo sample serves
as our simple mock galaxy catalog.  The ${\rm CDM_{full}}$ curve uses
the same minimum $V_{c}$ threshold for catalog inclusion as the WDM
sample, so there are $\sim2.8\times$ more halos in the ${\rm
  CDM_{full}}$ sample.  We include it here to show that the precise
value of the minimum halo $V_{c}$ does not much affect CDM clustering.
The clustering of mass (particles) is nearly identical in the two
cosmologies at scales larger than 10kpc, though this is partly due to
it being a mass-pair-weighted clustering measure, which is dominated
by particle pairs in large halos -- WDM small-scale clustering has
been shown to be suppressed in volume-weighted clustering measures
such as the power spectrum \citep{Vieletal2012,Schneideretal2012}.

Satellites are much
more strongly clustered than field objects as a result of being packed
into the small virial volumes of their host halos (the ``1-halo''
clustering component) combined with the high bias of massive hosts
(the ``2-halo'' clustering component).  Thus, small differences in low
mass structure suppression of satellites versus the field could
have a large effect on overall halo clustering, which we examine next.

\subsection{Environments of WDM and CDM halos}
\label{subsec-end}

In this section, we examine whether the suppression of WDM halos has
any environmental dependence, in particular, whether it is the same
for satellites versus field halos.  We consider first the probability
that a halo is a satellite or a field member in WDM and CDM.  A
satellite in this case is taken to be any halo whose center lies
within the virial radius of a larger parent (field) halo.

\begin{figure}
  \includegraphics[type=pdf,ext=.pdf,read=.pdf,width=.47\textwidth]{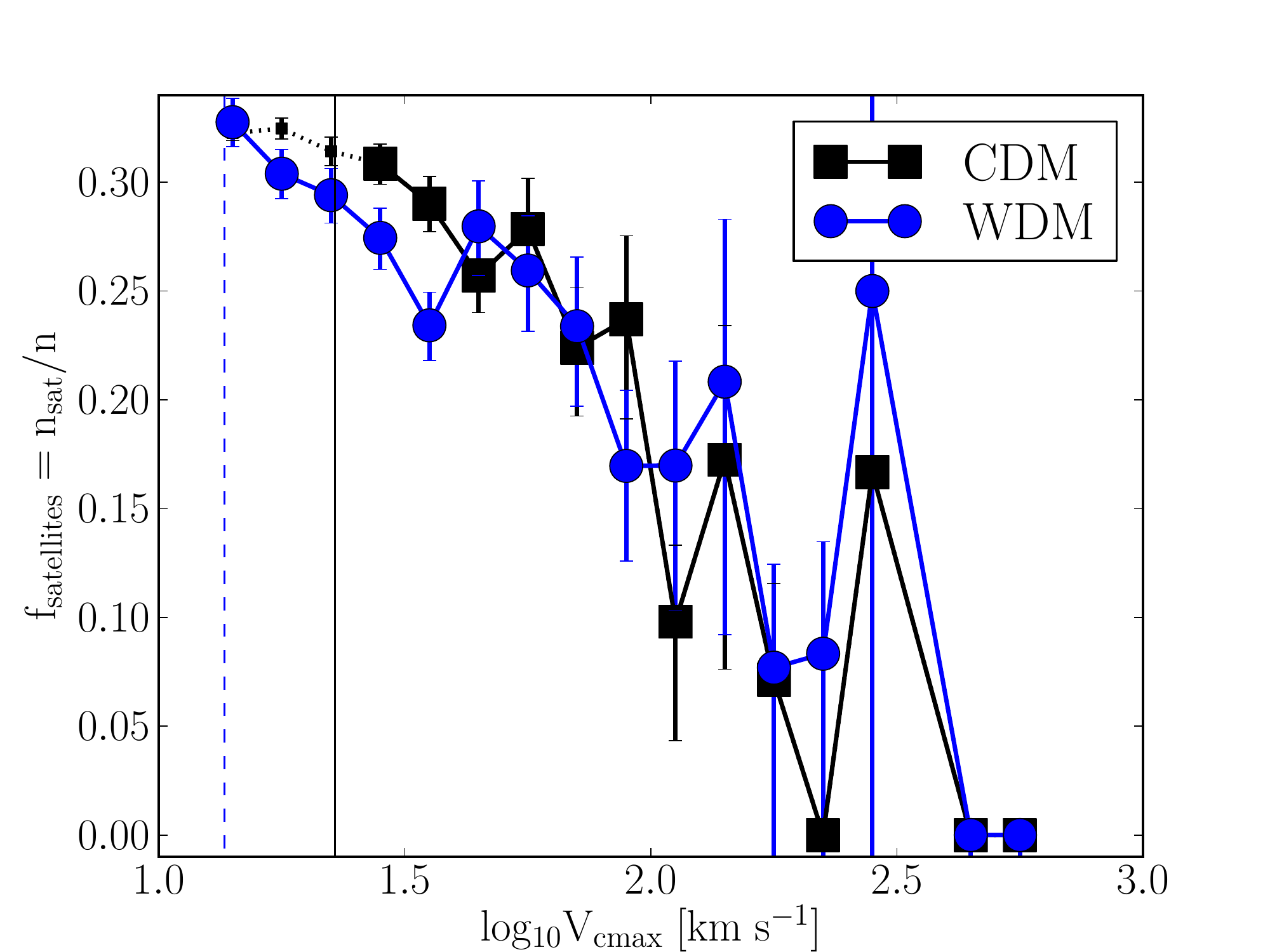}
  \caption{The probability that a mock galaxy in our catalogs is a satellite within a larger halo as a function of $V_{c}$.  A low $V_{c}$ mock galaxy is less likely to be a satellite in WDM cosmology than in CDM cosmology for the same $V_{c}$.  Vertical lines denote the minimum values for inclusion in the self abundance-matched pair of WDM and CDM mock galaxy catalogs.}
  \label{fig-fsats}
 \end{figure}

Given their similar overall clustering strengths, we would expect to
find similar satellite probability distribution functions (${\rm
  P_{sat}}$) in WDM and CDM in Fig. \ref{fig-fsats}.  However, the
mock galaxy catalog reveals that there is a higher probability to be a
satellite for small CDM halos than for WDM halos of equal $V_{c}$,
although this difference is only $\simlt 10\%$.  It is not obvious why
a WDM galaxy should be preferentially more likely to lie in the field
than to be a satellite compared to a CDM galaxy.  It may be that WDM
subhalos are more heavily stripped due to their lower concentrations.
In addition, even if the mass stripping is comparable for the two
cosmologies, as suggested by \citep{Elahietal2014}, it may cause a
larger decrease in $V_{c}$ for WDM halos because their lower
concentrations put their peak $V_{c}$ at larger radii.  The CDM
enhancement of ${\rm P_{sat}}$ for this mock galaxy catalog does {\it
  not} lead to enhancement in the CDM correlation function because
once the different minimum $V_{c}$ are considered, the WDM and CDM
samples have a similar total fraction of satellites.

\begin{figure}
  \includegraphics[type=pdf,ext=.pdf,read=.pdf,width=.52\textwidth]{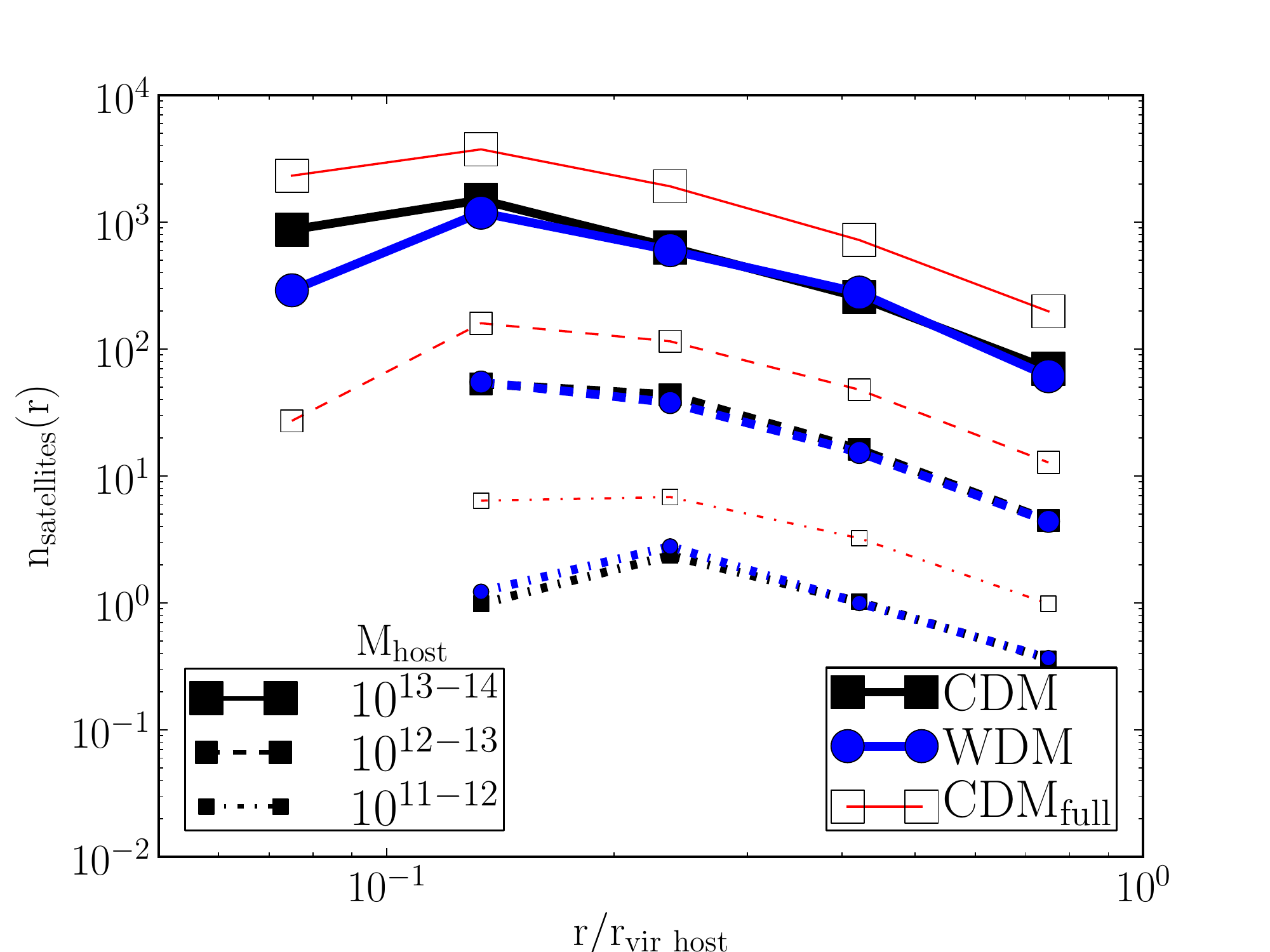}
  \caption{The stacked density profiles of the radial distribution of
    the population of $V_{c}$-selected satellites within their
    hosts. The only significant difference between CDM and WDM, once normalized by abundance, is that there are somewhat fewer WDM satellites near the the centers of the most massive
    halos, likely due to increased tidal stripping of the lower
    concentration WDM halos.  The ${\rm CDM_{full}}$ curves use the same minimum $V_{c}$ threshold as the WDM catalog, illustrating the higher satellite numbers when CDM is not abundance-matched to WDM. }
  \label{fig-stacked}
 \end{figure}

We now examine the host halo mass dependence of the radial profiles of
CDM versus WDM satellites.  The stacked radial number density profiles
of satellites within their host dark matter halos are shown in
Fig. \ref{fig-stacked} for three host halo mass ranges; as a reminder,
the abundance-matched construction CDM and WDM catalog removes the
zero order effect that there are fewer WDM halos.  The main difference
between the CDM and WDM satellite populations is that the central
regions of the largest halos, deficient of satellites in both
cosmologies, are somewhat more deficient in WDM.  This appears to
confirm that the enhanced tendency for CDM halos to be satellites in
$V_{c}$-selected mock galaxies (Fig. \ref{fig-fsats}) is due to
preferentially reducing $V_c$ by tidal stripping or destruction in the
central regions of massive halos, \ie clusters and large groups.

\begin{figure}
  \includegraphics[type=pdf,ext=.pdf,read=.pdf,width=.52\textwidth]{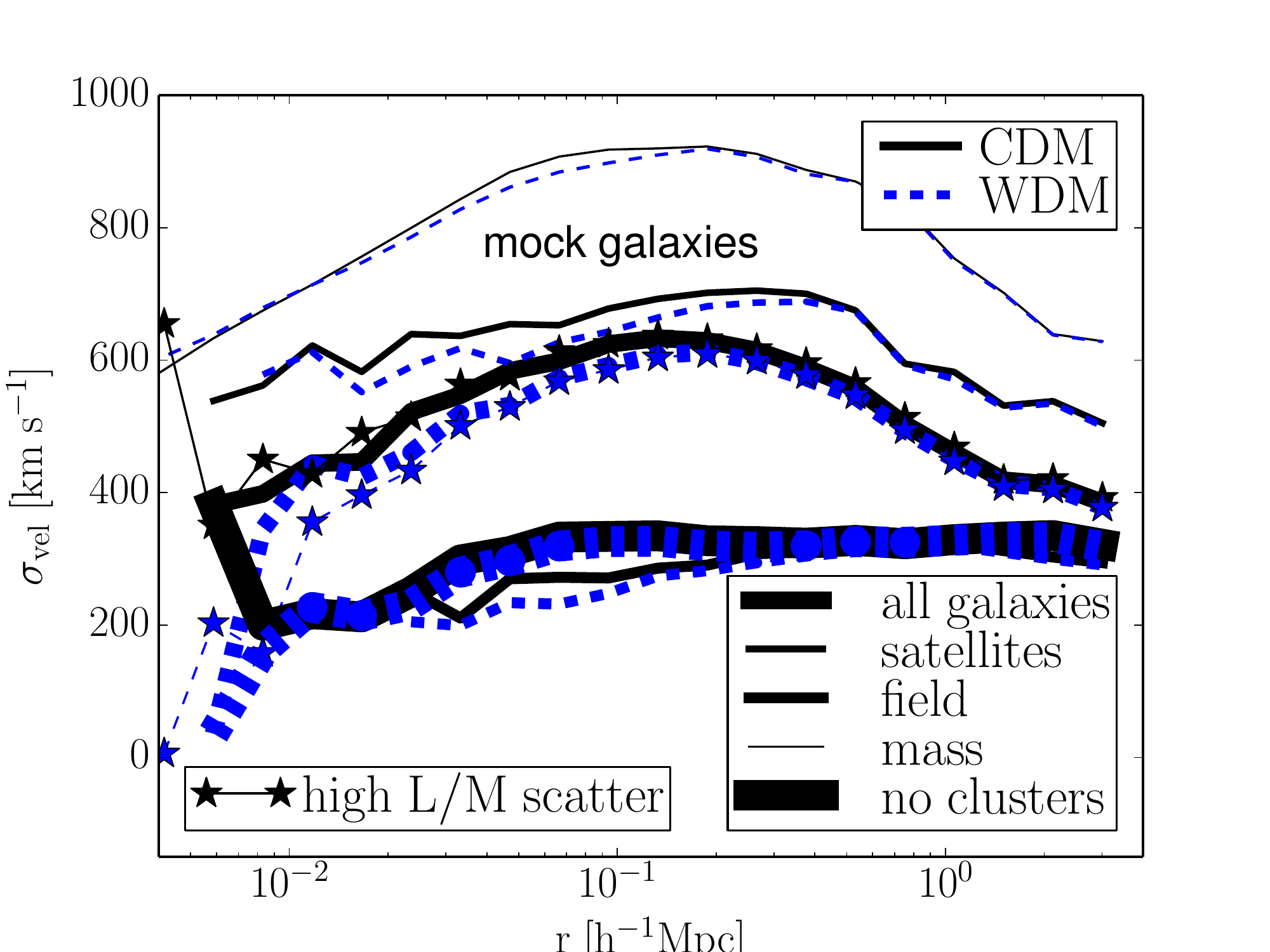}
\caption{The pairwise velocity dispersion, $\sigma_{vel}(r)$, of our mock galaxy catalog.  The reduced WDM velocities in the full sample (``all galaxies'') is mainly due to the central suppression of WDM satellites in the largest clusters ($>10^{13}\Msol$). Once those cluster members are excluded (thickest lines), the pairwise kinematics of WDM and CDM halos and mock galaxies are indistinguishable.}
  \label{fig-corrvel}
 \end{figure}

The similar spatial clustering and satellite to field ratios of halos
in the WDM cosmology imply that WDM halos should have similar peculiar
velocities.  Indeed, we have examined the mock galaxy halo pairwise
velocity dispersion, $\sigma_{vel}(r)$, defined as:
\begin{equation}
\sigma_{vel}(r)=\sum V_{\rm rel}^2/N_{\rm pairs},
\end{equation}
and we find indistinguishable values (within our uncertainty levels) for CDM and WDM, once the satellites of the largest clusters are excluded (Fig. \ref{fig-corrvel}).  The suppression of WDM satellites near cluster centers (Fig. \ref{fig-stacked}) combined with the disproportionately large effect of clusters in the pair-weighted $\sigma_{vel}(r)$ causes some suppression of WDM kinematics in the complete sample.  It is not obvious whether this effect could be detected given the other complex environmental effects that occur in clusters.  Any effects of the reduced halo concentrations of WDM halos on pairwise kinematics of their satellites are negligible because halo small enough to have altered concentrations \citep{Schneider2015} are also too small to host significant numbers of satellites.

\subsection{Why is WDM clustering so similar?}
\label{sec-discuss}

We have shown that WDM halos in the dwarf galaxy mass range are
similarly clustered in WDM and CDM because the suppression of WDM
halos is largely independent of environment.  Naively, it is perhaps
surprising that suppressing small-scale structure in WDM has little
effect on the clustering strength of small objects.  One might expect
WDM satellites to be much more easily tidally disrupted due to their
lower concentrations, which would lower their clustering strength.
One might also expect low mass WDM halos to be ``rare'' objects whose
formation might be enhanced by lying in a large scale
over-density, having the opposing effect of increasing their
clustering strength.  Regarding the homogeneity of WDM halo formation,
WDM suppression acts on low mass halos, which are relatively unbiased
with respect to the matter density field in CDM
\citep[\eg][]{Bondetal1991,ShethTormen1999,SeljakWarren2004,Tinkeretal2010}
and also in WDM \citep{SmithMarkovic2011}.  Hence, halo formation in
WDM is suppressed with nearly equal probability independent of larger
scale environment, (reflected by Fig. \ref{fig-corr}).  A thorough
theoretical description of WDM versus CDM halos bias is put forth by
\citet{SmithMarkovic2011} using a modified halo model of clustering
that accounts for the suppression of low mass WDM halos and also
includes the effects of WDM substructure suppression
\citep{Dunstanetal2011} on bias.

\begin{figure}
  \includegraphics[type=pdf,ext=.pdf,read=.pdf,width=.52\textwidth]{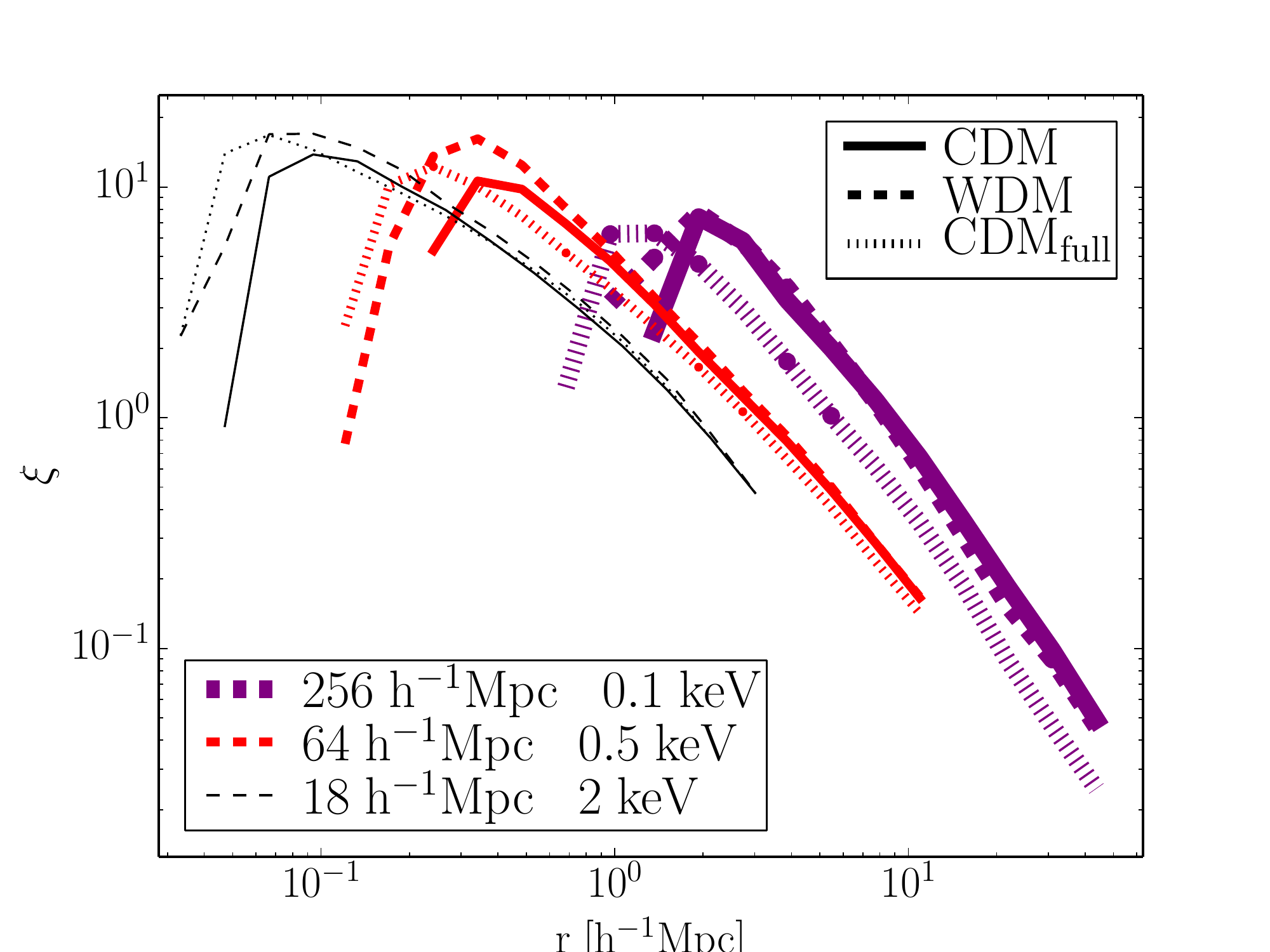}
  \caption{The correlation function of FoF halos for several different
    simulations of different box sizes and different thermal relic
    particle masses, with catalogs selected by mass.  When identical
    halo mass ranges are compared, as for `${\rm CDM_{full}}$' and
    WDM, clustering is very different with cosmology, primarily
    because the WDM sample has a flatter mass function, and hence, a
    higher average halo mass.  However, when samples are self
    abundance-matched, as for `CDM' and WDM, clustering is very
    similar.  Legend lists box length and WDM particle mass.  The
    2$keV$ run is used for the mock galaxies throughout this paper.  }
  \label{fig-fofmassboxes}
 \end{figure}

If we consider a more extreme (warmer) WDM cosmology with a
free-streaming cutoff near or above $M_*$, defined as the mass of a
$1\sigma$ over-density, $\simeq4.5\times10^{12}\Msol$ for CDM, halo bias can be significantly affected.  In
Fig. \ref{fig-fofmassboxes}, we show the bias of FoF halos for some
smaller WDM particle masses within larger simulation boxes (to capture
the larger free-streaming halo masses) taken from
\citet{Schneideretal2013} with the addition of a 0.1$keV$ WDM run,
each assuming a WDM mass resolution limit given by
Eqn.\ref{eqn-mfrag}.  If the WDM and CDM catalog pair is constructed
using identical minimum mass thresholds, the WDM sample is
significantly biased with respect to CDM -- the ``${\rm CDM_{full}}$''
and ``WDM'' lines -- as predicted by \citep{SmithMarkovic2011} and
shown in simulations by \citep{Schneideretal2012}.  This effect is
largest in the 0.1$keV$ run where the the half-mode WDM suppression
mass is $\sim10M_*$, and we sample to masses well below $M_*$.
However, when samples are abundance-matched to each other -- the CDM
and WDM lines -- there is little to no difference in clustering within
our $\sim 10\%$ uncertainties; we verified that this is also true for
FoF halos selected by $V_c$.  Loosely, this similar clustering
behavior occurs because the integral of $N(m)b(m)dm$ is constant when
the halo bias, $b(m) \simeq 1$.  A minor difference between the warmer
WDM models and their matching CDM catalogs is that the WDM halo pairs
extend farther into the ``exclusion scale'', approximately twice the
radius of a halo, due to their lower mass selection criteria and
corresponding smaller radii.


In summary, although WDM halos with a very warm particle are expected
to be more strongly biased than CDM halos, for plausible WDM particle
masses of $2~keV$, this WDM bias is very weak.  The weakness of WDM
clustering enhancement is due to the fact that the WDM halo
suppression scale is below $M_*$, and low mass halos are a reasonably
good tracer of mass.  Thus, suppressing halo formation at these
scales, has little effect on halo clustering, in line with theoretical
expectations.  However, there are implications for the effects on
clustering due to neutrinos and mixed dark matter models where the
streaming mass scale of the warm or hot component is large.

\section{WDM voids and neighbors}
\label{sec-voidsneighbors}

We discussed in \S~\ref{sec-introduction} that there is some
controversy as to whether the distribution of void sizes and the void
galaxy population are in agreement with CDM predictions.  The
emptiness of voids has been interpreted as a ``missing void dwarf
galaxy problem'' -- essentially the ``missing satellite'' problem in
low density regions.  And there also is the appearance that observed
voids are too large compared to CDM simulations when mock catalogs are
matched to the same minimum $V_c$.  These problems might be resolved
if low mass halos do not form many stars, but that solution would seem
to require large scatter between halo $V_c$ and galaxy luminosity to
avoid the ``too big to fail'' problem in the field.  In this section,
we explore whether the ``warmth'' of the dark matter power spectrum
might be probed by the distribution of small voids delineated by dwarf
galaxy halos, including the effects of scatter in galaxy luminosity to
halo mass.  We focus on void size statistics and nearest neighbor
statistics for mock dwarf galaxies.

\begin{figure}
  \includegraphics[type=pdf,ext=.pdf,read=.pdf,width=.47\textwidth]
{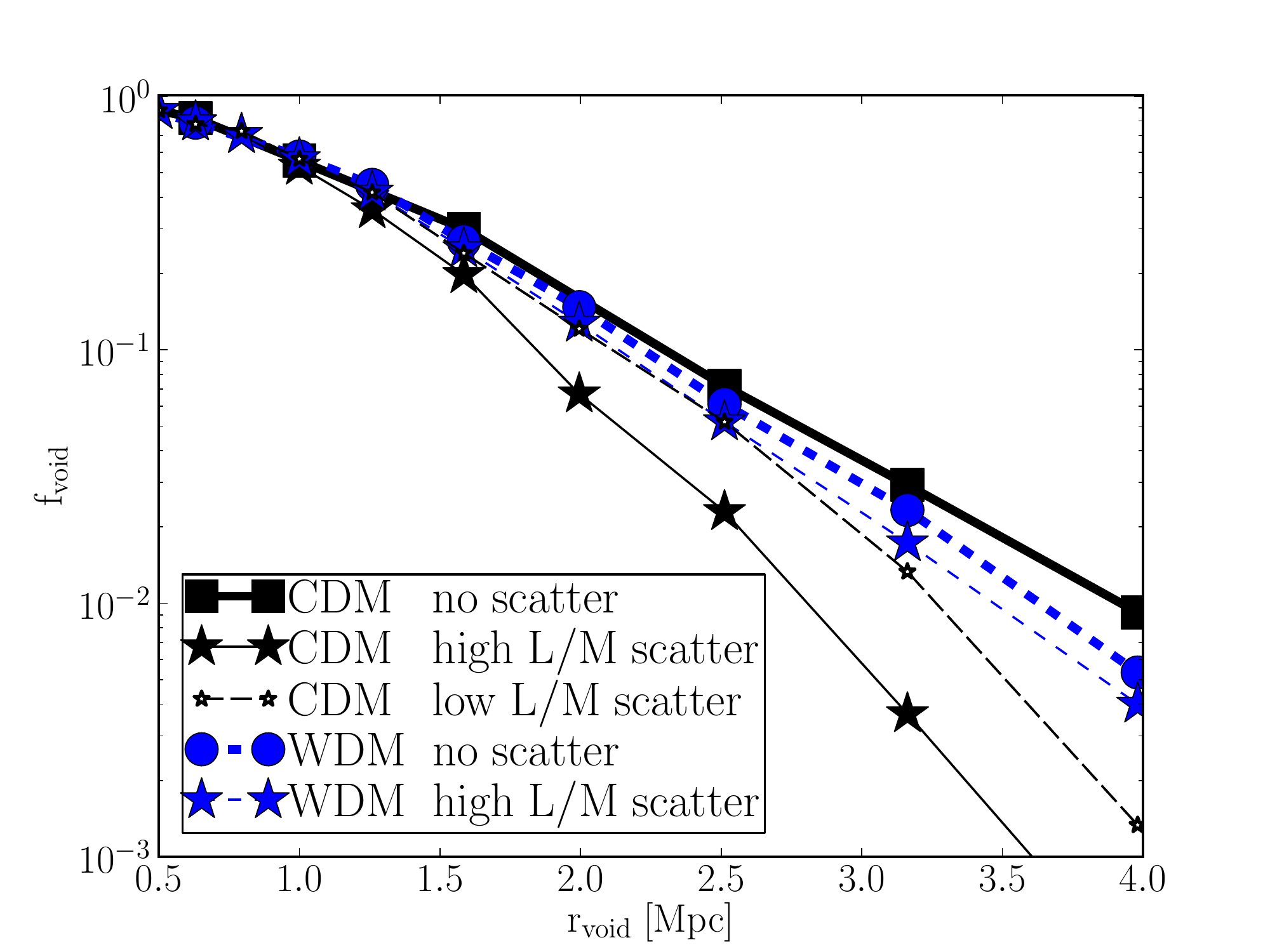}
  \caption{The void volume fraction for the WDM and CDM mock galaxy
    catalogs, determined by the fraction of randomly-placed spheres
    that contain zero galaxies.  Stars show the effect of a scatter
    between galaxy luminosity and halo $V_c$ (described in text),
    which reduces the numbers of CDM voids.}
  \label{fig-empty}
 \end{figure}

\begin{figure}
  \includegraphics[type=pdf,ext=.pdf,read=.pdf,width=.47\textwidth]
{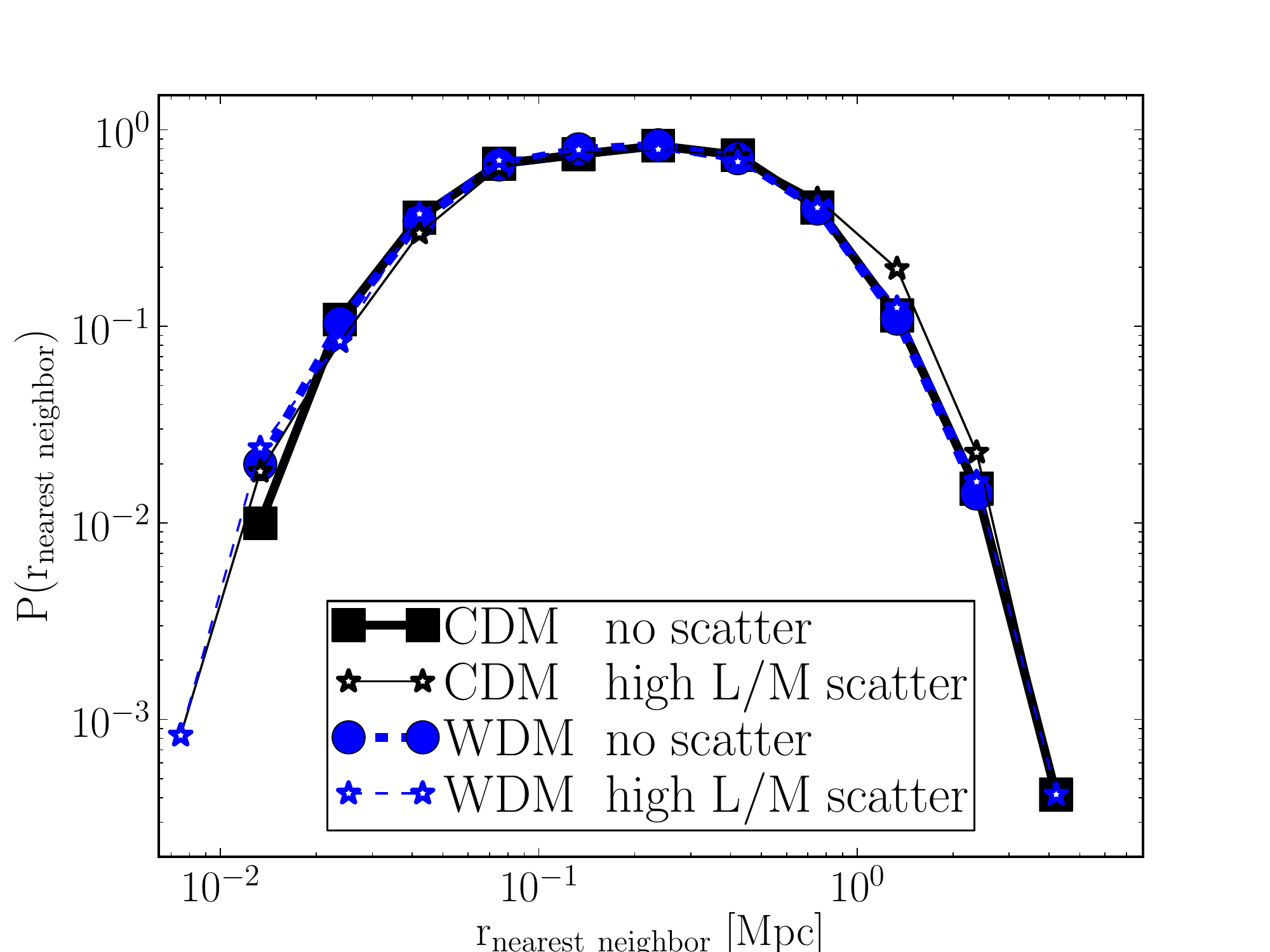}
  \caption{The probability distribution function of nearest neighbor
    distances for the WDM and CDM mock catalogs are also very
    similar to each other.  Symbols are identical to
    Fig. \ref{fig-empty}.}
  \label{fig-nn}
 \end{figure}

We show in Fig. \ref{fig-empty} that the void volume fraction, ${\rm
  f_{void}}$, is very similar in WDM and CDM in self abundance-matched
mock galaxy catalogs, assuming zero scatter between halo $V_c$ and
luminosity.  ${\rm f_{void}}$ is defined as the fraction of
randomly-placed spheres that contain no mock galaxies -- \ie the
fraction of the total volume occupied by empty (spherical) voids.
Note that the sensitivity of ${\rm f_{void}}$ to the galaxy number
density is removed since our CDM and WDM catalogs are self
abundance-matched.  The probability distribution function of nearest
neighbor galaxies is also very similar for the CDM and WDM catalog
when no scatter in galaxy to halo properties is assumed, shown in
Fig. \ref{fig-nn}.  This indicates that the fraction of isolated
galaxies and close pair galaxies does not depend on WDM versus CDM
cosmology.  In this sense, WDM voids are not larger or ``emptier'' of
halos than CDM, as is sometimes loosely stated, but are nearly
identical if one links galaxy properties to halo properties by simple
abundance matching.  In contrast to this invariance of void properties when defined by galactic halos, when defined by the mass distribution, small WDM voids are suppressed and WDM void profiles are flattened relative to CDM \citep[\eg][]{Yangetal2015}.

Imposing a scatter between galaxy luminosity and halo $V_c$ can have a
significant effect on void and neighbor statistics.  Recent
observations have shown that the scatter between luminosity and
circular velocity grows with decreasing $V_{c}$ for dwarf galaxies
\citep[\eg][]{Gehaetal2006}.  We consider the potential qualitative
effects of significant scatter between halo mass and luminosity on
${\rm f_{void}}$ and the nearest neighbor PDF.  We allow luminosity to
scatter by up to 2.5 magnitudes (a factor of ten scatter in mass to
light ratio) for low mass halos and limit the scatter to 0.4
magnitudes for halos larger than ${\rm 80 km~s^{-1}}$, following this
formula to describe the rms scatter in galaxy magnitude:
$\sigma_{Mag}(V_c) = 0.4 + (V_c/80km~s^{-1})^{-2}$, where
$\sigma_{Mag}(V_c)=2.5$ for $V_c<{\rm 55 km~s^{-1}}$.  This denotes
our `high $L/M$ scatter' example.  The somewhat arbitrary scatter is
chosen to be large enough to demonstrate its effect; it is larger than
that inferred from the stellar mass $V_c$ relation for dwarf galaxies
of \citep{Gehaetal2006}.  However, we note that the same study shows
that luminosity to mass ($L/M$) scatter appears to increase for
smaller galaxies.  In addition, local group dwarfs show evidence of
extremely large $L/M$ scatter \citep{Strigarietal2008}.  We also
consider a `low $L/M$ scatter' example, which imposes a constant
scatter of 1 magnitude below ${\rm 80~km~s^{-1}}$.  We see in
Fig. \ref{fig-corr} that $L/M$ scatter has little effect on halo
correlation functions (stars).  However, in Fig. \ref{fig-empty}, the
void volume fraction (${\rm f_{void}}$) becomes significantly lower
with increasing void size for CDM voids because some low mass void
halos, of which CDM contains more, can be populated by relatively
bright galaxies.  Moreover, the shape of the ${\rm f_{void}(r)}$
changes relative to the fiducial CDM catalog, which has no scatter in
$L/M$.  The effect is much smaller for WDM due to the lower numbers of
small halos available to scatter up in luminosity and populate the
voids.  Fig.\ref{fig-nn} shows that the effect of $L/M$ scatter on
nearest neighbor statistics is perhaps also important.  There is a
$\sim50\%$ increase in isolated galaxies with nearest neighbors of
${\rm \sim2Mpc}$, corresponding to halos in otherwise empty voids.

We chose to use a spherical void definition for its simplicity.  Because spheres will not usually align well with the irregular boundaries of voids, our spherical voids likely are sensitive primarily to effects in the central regions of voids.  Had we instead chosen a void definition that more accurately traces the shapes of individual voids, we might expect to be more more sensitive to any effects near the edge of voids

To summarize this section, the void and neighbor statistics that we
have considered would be unlikely to be useful as a means to
distinguish WDM from CDM if the scatter between halo $V_c$ and galaxy
luminosity is small.  However, if small halos have a large scatter in
galaxy to halo properties, completely empty CDM voids would occupy a
lower total volume due to a small increase in isolated galaxies
relative to WDM.

\section{Implications for the cosmological galaxy population}
\label{sec-galaxies}

The cosmic web of halos is remarkably similar in a CDM and a WDM
universe once the reduced number of WDM halos is accounted for by halo
abundance matching.  The similar clustering properties of WDM and CDM
halos imply that it will be difficult to use dwarf galaxy clustering
as a test of WDM versus CDM cosmology unless the scatter in halo to
galaxy properties is rather large.  In that case, some differences in
the distribution of galaxies emerge.  In particular, CDM voids would
be less empty than WDM voids because some low mass CDM halos would
scatter up to high luminosities.  Curiously, \citet{PeeblesNusser2010}
cite the existence of relatively bright but isolated galaxies in the
local volume as a potential CDM problem because massive halos strongly
prefer to lie in the walls, filaments, and knots that bound void
volumes.  Our results imply, conversely, that within a CDM universe
that also has large halo mass to luminosity scatter, there is the
potential for a population of relatively isolated and bright galaxies,
possibly alleviating any such problem.

If in fact we live in a CDM universe, a large scatter between galaxy
luminosity and halo mass is suggested by observations.  One example is
the previously mentioned Tully-Fisher relation for dwarfs, which
appears to have a scatter of order unity in luminosity and in stellar
mass for halos in the $20-50 ~ km~s^{-1}$ range
\citep[\eg][]{Gehaetal2006}.  Considering luminosities even fainter
than where the Tully-Fisher relation is well-constrained, the common
300 parsec dynamical mass scale of $10^7\Msol$ over $\sim 5$ orders of
magnitude in luminosity for Milky Way satellites
\citep{Strigarietal2008} implies a very large scatter in dwarf galaxy
luminosity to halo mass.  CDM can remain viable if the proposed
baryonic solutions to CDM problems can flatten the velocity
function and the luminosity function in accord with the steep halo
mass function, issues that are topics of numerous studies.  It would
not be surprising if these baryonic processes have some stochastic
elements that lead to scatter between galaxy and halo properties.
These issues suggest that the statistics of void sizes and isolated
galaxies could be useful not just to indicate how warm our cosmology
might be, but also to learn how galaxies populate halos.  Because the
astrophysics of galaxy formation is very difficult to model directly,
void statistics could thus provide useful clues about low mass galaxy
formation.

Whether the scatter between galaxy luminosity and halo mass is large
enough for galaxies bright enough to populate a sufficiently large
volume that galaxy surveys may distinguish between WDM and CDM from
void statistics is an open question.  With sufficient statistics, one
could determine whether the distribution of galaxies within voids and
their boundaries is consistent with that expected from halos in CDM
simulations
\citep[\eg][]{TinkerConroy2009,Ricciardellietal2014,Hamausetal2014b}.
A potential difficulty to using small-scale void statistics as a probe
of the matter power spectrum is that we have so far assumed that the
$L/M$ scatter depends only upon halo mass or $V_c$, \ie that galaxy
properties are independent of local environment.  Environmental
dependence of the galaxy-halo relation could be important.  For
example, voids may have a different UV heating history due to
inhomogeneous reionization or other astrophysical processes
\citep{SobacchiMesinger2013}, which could affect gas infall, star
formation and galaxy-halo relations, perhaps washing away any effects
on void statistics of the WDM versus CDM particle.  In addition, later
halo formation in voids makes them more prone to suppression of gas
infall and star formation by UV photo-heating \citep{Hoeftetal2006}.
For success, it may be necessary to obtain constraints on the scatter
between galaxy and halo properties in different environments.
Nonetheless, it is potentially significant and worthy of future
exploration that the dark matter particle type may be imprinted on the
cosmic web of galaxies.  Forthcoming surveys results, the Large Synoptic Survey Telescope (LSST) in
particular, will greatly improve our 3D positional mapping of dwarf
galaxies and small voids in the local volume by allowing distance
measures from stellar based techniques.

We briefly consider some possible implications that the different host halo mass ranges could have upon the galaxy populations of the two abundance-matched cosmologies.  In our mock galaxy catalog, without $L/V_c$ scatter, galaxies populate all WDM halos with ${\rm V_c > 13.7 km ~s^{-1}}$, but the host halo population for CDM galaxies extends only to ${\rm V_c > 22.9 km ~s^{-1}}$, corresponding to a factor of $\sim 3$ in halo mass.  Significant differences may thus be present in the properties of galaxies of similar luminosity due to the differing halo masses in the two cosmologies.  Low mass galaxies in the WDM cosmology would have significantly delayed star formation histories relative to CDM galaxies \citep{Caluraetal2014, Sitwelletal2014, Dayaletal2015, Maioetal2015, Governatoetal2015}.  This delay in star formation is partly due to the delay in the formation of WDM halos at fixed mass.  In addition, CDM halos, due to their higher masses, will more quickly surpass the mass threshold where baryon cooling by atomic transitions, and thus galaxy formation, is expected to be efficient (${\rm \sim 10 ~km ~s^{-1}}$, ${\rm \sim ~10^4K}$).  However, if we instead live in a CDM universe with large $L/V_c$ scatter, then some CDM galaxies should lie in lower mass halos.  This would imply a large scatter in star formation histories at fixed galaxy luminosities with respect to the CDM low $L/V_c$ scatter or WDM models.  In this case, star formation may be delayed in some CDM galaxies with respect to the WDM case.  These differences in galaxy properties might be detectable in local dwarfs or in the high redshift galaxy population in the future as observations and our understanding of galaxy formation improve.

\section{summary}
\label{sec-summary}

We explore dwarf galaxy halos in a pair of warm dark matter (WDM) and
cold dark matter (CDM) cosmological simulations that differ only in
that the WDM initial transfer function is suppressed on small scales
to approximate the effects of a $2~keV$ thermal dark matter relic
particle on the matter fluctuation spectrum.  Abundance-matched CDM
and WDM catalogs allow a fair comparison between a WDM cosmology and a
CDM cosmology where low mass galaxy formation is truncated below some
halo mass (or virial velocity) threshold.  In such abundance-matched
CDM-WDM pairs of mock galaxy catalogs, if galaxy properties (\eg
luminosity) monotonically trace halo circular velocity, the features
of the cosmic web that we consider -- the pair correlation function,
pair kinematics, the void volume fraction, and the isolated galaxy
fraction -- are very similar in WDM and CDM.  This reflects the
similar distributions of WDM and CDM halos.  However, in the case of a
CDM universe with large scatter in the relation between galaxy and low
mass halo properties, a prospect that is well-motivated to consider,
CDM would contain an increased population of relatively isolated and
bright galaxies.  Such a population of void galaxies would be
difficult to explain in a WDM universe.


\section{acknowledgments}
The simulations were performed on Rosa at the Swiss National
Supercomputing Center (CSCS), and the zbox3 and Schr\"odinger
supercomputers at the University of Zurich. RES acknowledges support
from a Marie Curie Reintegration Grant and the Alexander von Humboldt
Foundation.  We thank George Lake, Jonathan Coles, Michael Busha, and
Donnino Anderhalden for insightful discussions.  We are grateful to
the Centro de Ciencias de Benasque Pedro Pascual where some
preparation of this draft was done during the 2014 ``Modern
Cosmology'' workshop.  We are grateful for the constructive suggestions of the anonymous referee.



\bibliographystyle{mn2e}
\bibliography{refsDarren}


\begin{appendix} 

\section{Numerical issues ~~ or \\ How I learned to stop worrying and love the simulation}
\label{sec-strangelove}

This section addresses some of the caveats to this necessarily
imperfect numerical simulation.  It is possible, with sufficient
computational power, to achieve percent level accuracy for many
properties of gravity-only simulations due to their relative
simplicity \citep[\eg][]{Heitmannetal2010,Reedetal2013}.  Our
conservative effective halo resolution limit of 1000 particles avoids
the sensitivity of low mass halos to numerical problems such as mass
discreteness effects
\citep[\eg][]{Melott1990,JoyceMarcos2007,Joyceetal2009}, time-stepping
or force accuracy \citep[\eg][]{Reedetal2013}.  However, WDM
simulations present a particular challenge at the dwarf-galaxy scales
in which we are most interested.  The low amplitude of small-scale
cosmological density perturbations means that artificial perturbations
due to mass-discreteness or force errors are relatively larger with
respect to the real perturbations.

\citet{WangWhite2007}, and later, \citet{Schneideretal2012},
\citet{Anguloetal2013} and \citet{Lovelletal2014}, showed that below
some halo mass scale, the mass function of WDM simulations becomes
dominated by spurious structure, noting that filaments fragment into
small pieces separated uniformly by the mean particle separation.
This results in a striking ``beads on a string'' visual effect and is
clearly a numerical artifact, as confirmed by mass resolution
convergence tests.  By comparing the mass function in simulations with
varying mass resolution, they determined that the resolved WDM mass
scale increases very slowly with finer resolution, $M_{halo,min}
\propto M_{particle}^{1/3}$.  This convergence showed that WDM mass
function is relatively robust provided that one ignores all halos
below the resolved scale.  The resolved halo mass scale of CDM
simulations, conversely, appears to scale much better, and generally
appears to be proportional to particle mass
\citep[\eg][]{Jenkinsetal2001,Warrenetal2006,Reedetal2007b,Trentietal2010,Bhattacharyaetal2011}.

To mitigate the WDM resolution problem, we apply a spurious halo
removal that requires the initial condition Lagrangian regions of
particles that will end up in AHF halos be approximately spherical, a
test shown by \citet{Lovelletal2014} to significantly reduce
artificial WDM halos.  The method consists of computing the shape
parameters ($c<=b<=a$) of every proto-halo patch in the initial
conditions. Haloes with unusually elongated initial patches,
$c/a<0.2$, are marked as artificial and omitted from the sample.  See
\citet{Lovelletal2014} for a detailed explanation of the method.  
For technical reasons, we measure the initial conditions shape of the material present in each halo at $z=0$ rather than measuring the halo material at the epoch where the halo has reached half its maximum mass, as advocated by \citet{Lovelletal2014} -- this would allow satellite halo initial shapes to be captured before accretion into larger halos and subsequent stripping, which could erase the signal of spurious halo formation.
We note that our spurious halo removal identifies more field objects than
satellites as spurious, which may reflect that spurious halos are
formed from fragmentation of filaments in the field.  We note, however, that our implementation may be less effective at detecting spurious satellite halos that have been stripped of their outer layers, which might be expected to contribute
the most to asphericity of the initial patch.  
Our minimum halo $V_c$
threshold is high enough that this spurious halo removal does not
impact our results, but instead confirms that we are free from
significant contamination.

Fig. \ref{fig-mf} shows the FoF and AHF mass functions, which are
relatively similar to each other.  The spurious halo correction to our
AHF catalog has little effect on the mass function above our mass and
$V_c$ resolution limits.  Ignoring halos below our effective sample
$V_c$ resolution limit of one thousand particles, the FoF mass
function is consistent with the ``WDM prediction'' mass function of
\citep[][]{Schneideretal2013}, based on excursion set theory.  Our
minimum halo mass for inclusion in the catalog is chosen to be above
the mass where artificial halos dominate this model.  Our corrected
AHF mass function diverges from the WDM mass function fit
approximately at our fiducial $10^3$ particle minimum halo $V_c$ and
mass.  This implies that halos below our resolution limit are heavily
contaminated by spurious structures whereas halos more massive than
our resolution limit are relatively uncontaminated.  Importantly, the
mass scale below which the mass function upturns, indicating spurious
halos, is approximately equal for AHF halos (including subhalos) and
FoF halos.  In only one case, where high scatter between luminosity
and halo $V_c$ is assumed, do we allow halos below threshold
resolution to scatter into the catalog.  This does not pose an issue
because if some of the WDM halos are spurious in Fig. \ref{fig-empty}
and \ref{fig-nn}, the differences between CDM and WDM would decrease.
Instead the WDM high scatter sample is very similar to the WDM zero
scatter sample.

The halo shape exclusion criteria detects significant artificial
structure beginning at approximately 200 particles.  The departure
from the \citet[][]{Schneideretal2013} prediction, however, occurs
beginning at approximately 1000 particles.  This could indicate that
the halo shape test is incomplete at identifying artificial structure
or that this WDM mass function prediction is too low.

We note that the spurious upturn is also apparent in the uncorrected
circular velocity function of halos (bottom panel of
Fig. \ref{fig-mf}).  Our conservative minimum ${V_{c}}$ is matched to
the abundance of the mass-selected catalog, yielding a selection
criteria well above the spurious upturn in the circular velocity
function.  We have checked the effect of removing unbound particles
from FoF halos (\ie ``unbinding'') based on each halo potential --
most FoF halos with more than $\sim 100$ particles are self-bound
collections of particles, whether spurious or not.

Because the spurious halo mass range extends to masses well above the
inflection point of the FoF spurious halo upturn
\citep[][]{Schneideretal2013,Anguloetal2013,Lovelletal2014}, it is not
practical to choose a mass large enough to definitively avoid all
contamination.  At present, we cannot rule out the possibility that
some effects of spurious WDM halos persist at several times higher
masses than our 1000 particle spurious mass threshold.  Since
artificial halos live preferentially in denser regions
\citep{Schneideretal2013}, there is a possibility for our WDM sample
to have systematic biases in clustering properties.  However,
reassuringly, convergence occurs well below the mass scale where WDM
numbers are suppressed, implying that we have captured the important
physical effects of WDM above our resolution limits.  Moreover,
catalogs selected by minimum circular velocity, such as our mock
galaxy catalogs, converge more quickly than catalogs selected by mass.
Our results are converged to the $\sim 10\%$ level for different
choices of WDM minimum $V_c$ thresholds in our catalog, which
indicates that the mass-dependent WDM spurious halo fraction is not
important.

Finally, we note that for this study we have available only the
present-day $V_c$ for our halos.  Results of the SubHalo Abundance
Matching (SHAM) technique of mapping galaxy luminosity to halo masses
\citep[\eg][]{ValeOstriker2004,Kravtsovetal2004,Conroyetal2006},
suggest that the peak halo mass, or its mass prior to infall before
becoming a satellite, is more closely related to its present-day
luminosity
\citep[\eg][]{Conroyetal2006,ValeOstriker2006,Mosteretal2010,Rodriguez-Pueblaetal2012,Reddicketal2013}.

The issues discussed here provide some confidence that the halo mass
range we have chosen is sound with regards to the relative CDM versus
WDM clustering measures that we consider.  Further improvement in WDM
simulations may be possible with a new simulation technique by
\citet{Anguloetal2013} that is able to largely avoid forming spurious
WDM structure.

\subsection{Why do artificial halos form - a toy model}

We now describe a simple intuitive toy model for spurious halo
formation.  If one assumes that the WDM resolution scale is determined
entirely by filament fragmentation, one can reproduce the empirical
scaling resolution for the minimum resolved halo mass.  A filament
whose Lagrangian (pre-collapse) radius is equal to the free-streaming
scale will have little to no cosmological power on transverse scales
below this free-streaming scale.  However, the formation of the
filament by radial collapse of discrete masses will result in clumps
of particles.  If the filament is aligned with a grid axis, particles
will be in clumps along the filament with spacing equal to the initial
grid spacing; this is simply due to the initial grid geometry (though
with noise if a `glass-like' instead of a grid initial particle
distribution is used).  For filaments at other angles, the densities
of these filament particles will still be highly inhomogeneous,
implying that they will readily collapse to form halos.  In CDM, this
spurious fragmentation does not seem to occur because there is
sufficient small-scale power that particles clump together from real
cosmological perturbations.

One can estimate the mass scale at which this effect
becomes strong, substituting the half-mode mass scale for the free-streaming scale (see Eqn \ref{eqn-mhm}):
\begin{equation}
M_{\rm fragment} = \kappa \overline{\rho} \pi \lambda_{\rm hm}^2 l_{\rm grid},
\label{eqn-mfrag}
\end{equation}
where ${\rm l_{grid}=L_{box}N_{part}^{-1/3} }$ is the initial
inter-particle spacing.  The empirical mass-scaling factor, $\kappa
\approx 0.5$, is required to calibrate the spurious pre-collapse
Lagrangian radius, which is of order $\lambda_{\rm hm}$.  Our estimate
of $\kappa$ is determined by the mass at which the WDM mass function
begins to deviate upward from the \citet{Schneideretal2013} fit,
$M_{\rm fragment} \approx 500$.  It would not be surprising if
$\kappa$ depends weakly on cosmology or mass-resolution.

This intuitive toy model recovers the empirically determined
scaling of the effective WDM halo mass resolution
\begin{equation}
M_{\rm fragment} \propto M_{part}^{1/3}
\end{equation}
below which the WDM mass function becomes a steep power due to
spurious fragmentation \citep{WangWhite2007}.  Our toy model for
$M_{\rm fragment}$ is a good match (within $\sim2\times$) to the
artificial structure upturn for several simulations in the literature
\citep[\eg][]{Zavalaetal2009,Schneideretal2013,Anguloetal2013,Lovelletal2014}.
At scales significantly larger than $M_{\rm fragment}$, there should
be sufficient power to promote genuine longitudinal filament
fragmentation.  Our one thousand particle effective halo mass
resolution threshold is conservative, well above $M_{\rm fragment}
\approx 500$.  $M_{\rm fragment}$ is generally consistent with $M_{\rm
  lim}$ of \citep{WangWhite2007} though the relative difference varies
somewhat with free-streaming scale because we use the transfer
function suppression scale ($\lambda_{\rm hm}$) whereas they use the
power spectrum peak wave number to estimate a fragmentation scale.

\section{Mass-selected halos}
\label{sec-massselection}

Throughout the paper, we select halos by $V_c$ because it is less
vulnerable to the effects of tidal stripping and so should better
correlate with galaxy properties.  When one instead considers
mass-selected halos, some differences arise.  Mass selection is more
appropriate for studies that consider the effects of lensing
substructure.

\begin{figure}
\includegraphics[type=pdf,ext=.pdf,read=.pdf,width=.52\textwidth]{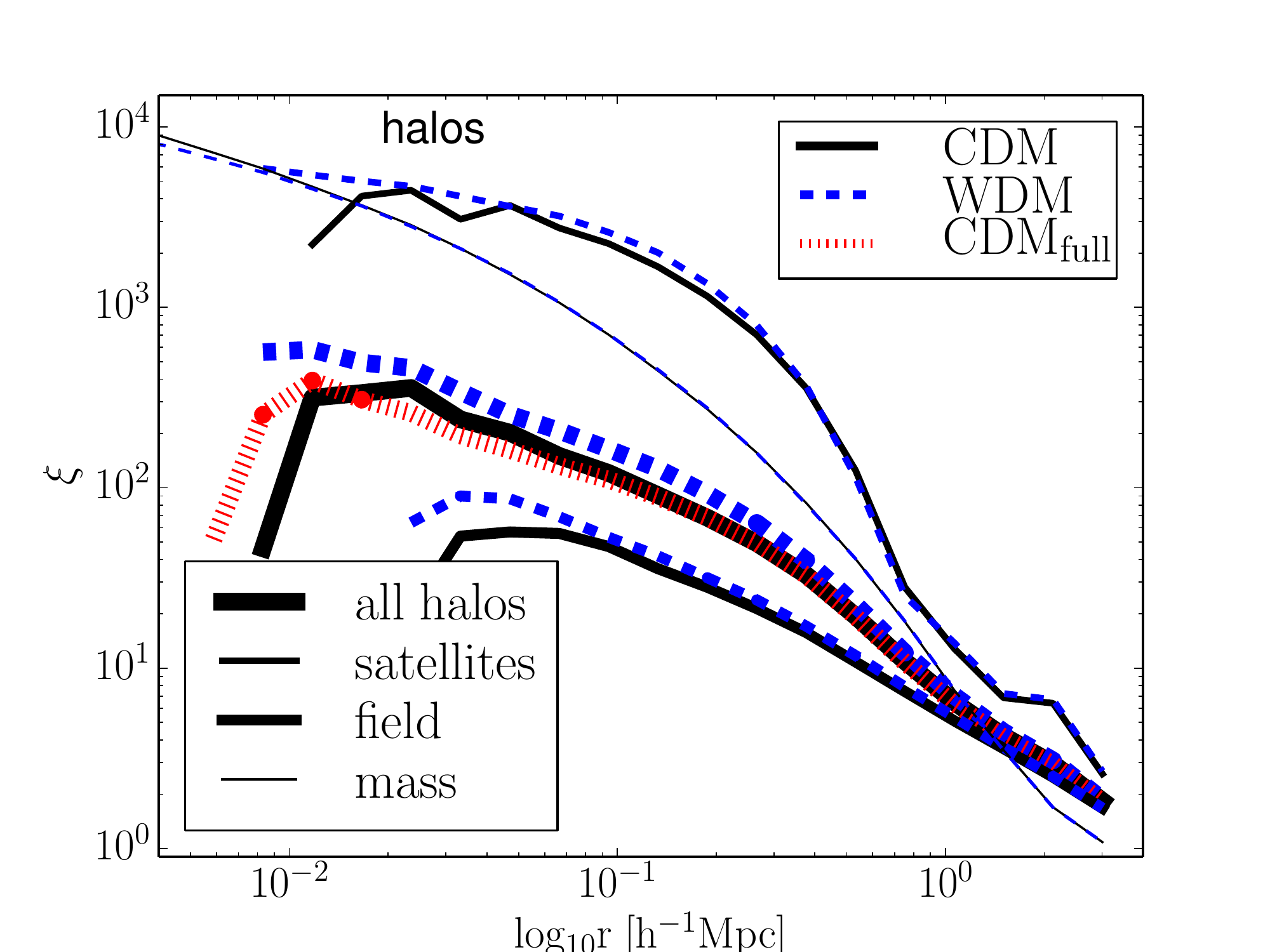}
  \caption{The correlation function of mass-selected halos more
    massive than $5.5 \times 10^{8}\Msol$ for WDM (1000 particles) and
    $1.7 \times 10^{9}\Msol$ for CDM, self abundance-matched.  WDM
    mass-selected are by a small significant amount, as opposed to
    $V_c$-selected halos wherein WDM and CDM clustering is
    approximately equal.  CDM$_{\rm full}$ uses the same mass
    threshold as the WDM sample, showing that the selection criteria
    has little effect on CDM halo clustering.  }
  \label{fig-masscorr}
 \end{figure}

\begin{figure}
\includegraphics[type=pdf,ext=.pdf,read=.pdf,width=.47\textwidth]{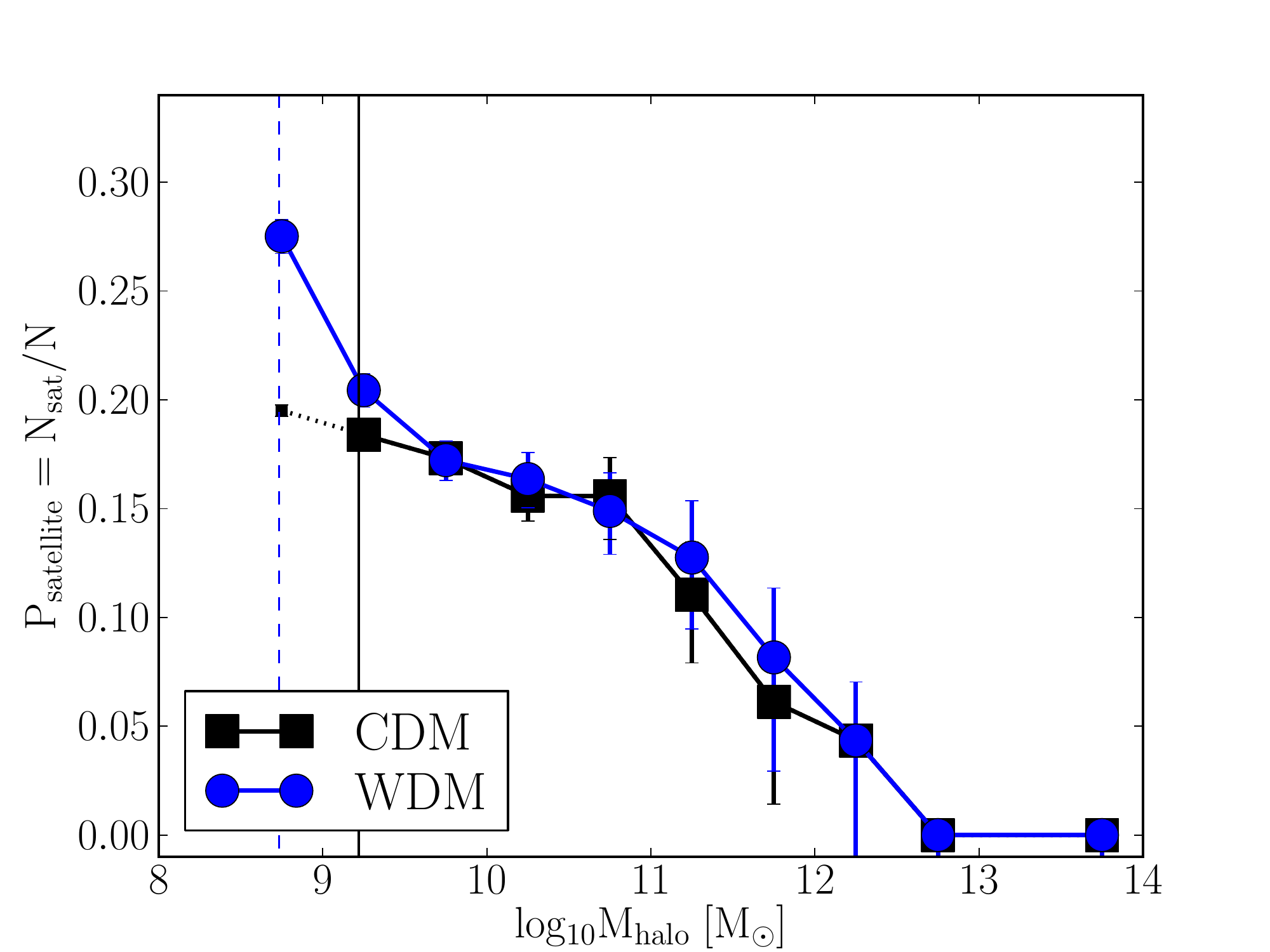}
  \caption{The probability that a mock galaxy is a satellite of a
    larger halo as a function of mock galaxy halo mass.  A low mass
    halo is more likely to be a satellite in WDM cosmology than a CDM
    halo of the same mass, contrary to the case for $V_{c}$-selected
    halos (Fig.\ref{fig-fsats}).  Vertical lines denote the minimum
    values for inclusion for the self abundance-matched pair of WDM
    and CDM mock galaxy catalogs.  The likely explanation is that many
    low mass WDM satellites are stripped higher mass objects (see
    text).}
  \label{fig-massfsats}
 \end{figure}

Fig. \ref{fig-masscorr} (top panel) compares a mass-selected CDM and
WDM halo catalog, matched to each other by abundance, which is set by
including all WDM halos with more than 1000 particles, or $5.5 \times
10^{8}\Msol$, leading to a CDM abundance-matched mass limit to $1.7
\times 10^{8}\Msol$.  The mass-selected WDM halos have a weak
($\sim50\%$) clustering enhancement on scales below a few hundred
$\kpc$ -- approximately the virial radius of $M_*$ halos.  By
contrast, when $V_{c}$-selection is used for the catalogs, clustering
was nearly identical (Fig. \ref{fig-corr}).  We see in
Fig. \ref{fig-massfsats} that low mass WDM halos are more likely than
CDM halos to be satellites inside larger halos, which should account
for the overall clustering difference.  One possible explanation for
the enhancement of WDM mass-clustering is that because of the large
amount of mass stripping satellites undergo, many satellites were much
more massive in the past in both cosmologies.  The low mass satellite
WDM population then consists of both relatively unstripped low mass
halos that have been accreted and heavily stripped halos that were
more massive upon accretion.  The stripped massive halo population is
proportionally much larger in WDM than CDM because the WDM mass
function is relatively flat.  This does not imply more small
substructures in absolute terms for WDM; low mass halo numbers are
reduced in all environments, but more so in the field.

\end{appendix}

\label{lastpage}
\end{document}